\documentclass[a4paper]{amsart}
\usepackage{RR}
\RRNo{6331}
\usepackage{hyperref}
\RRdate{October 2007}

\newcommand{\GF}{\mathrm{GF}}
\usepackage{url}
\usepackage{epic,eepic}

\newtheorem{theorem}{Theorem}[section]
\newtheorem{lemma}[theorem]{Lemma}

\newtheorem{hypothesis}[theorem]{Hypothesis}

\def\GCD{\mathrm{GCD}}
\newcommand{\Otilde}{\makebox{$\widetilde O$}}

\def\endcomma{,} 
\def\enddot{.}   

\RRauthor{
Richard P.\ Brent\thanks{Mathematical Sciences Institute, Australian National
  University, Canberra, ACT~0200, Australia, \url{Fq8@rpbrent.com}}%
\and%
Paul Zimmermann\thanks{Centre de Recherche INRIA Nancy - Grand Est,
615 rue du Jardin Botanique, 54600 Villers-l\`es-Nancy, France,
\url{Paul.Zimmermann@inria.fr}}
}
\authorhead{Brent \& Zimmermann}
\RRtitle{Un algorithme multi-\'etage
  pour la factorisation en degr\'es distincts
         de polyn\^omes}
\RRetitle{A Multi-level Blocking Distinct Degree\\ Factorization Algorithm}
\titlehead{A Multi-level Blocking Distinct Degree Factorization Algorithm}
\RRresume{Nous proposons un nouvel algorithme pour la factorisation en
degr\'es distincts d'un polyn\^ome $P(x)$ sur $\GF(2)$, via une strat\'egie
multi-\'etage.
Le niveau sup\'erieur remplace des calculs de pgcd par des multiplications,
comme sugg\'er\'e par Pollard (1975), von~zur Gathen et Shoup (1992), et
d'autres auteurs.
L'originalit\'e de notre approche tient dans un niveau inf\'erieur, qui
remplace des multiplications par des carr\'es, ce qui acc\'el\`ere le calcul
de certains \emph{polyn\^omes intervalles} sur $\GF(2)[x]/P(x)$,
quand $P(x)$ est creux.

Comme application nous exhibons un algorithme rapide cherchant tous les
trin\^omes irr\'eductibles $x^r + x^s + 1$ de degr\'e $r$ sur $\GF(2)$,
tout en produisant un certificat qui peut \^etre v\'erifi\'e plus
rapidement.
Les algorithmes na\"{\i}fs co\^utent $O(r^2)$ par trin\^ome, soit $O(r^3)$
pour tous les trin\^omes de degr\'e $r$.
Sous une hypoth\`ese naturelle sur la distribution des facteurs de trin\^omes,
le nouvel algorithme a une complexit\'e
$O(r^2 (\log r)^{3/2}(\log\log r)^{1/2})$ pour tester tous les trin\^omes
de degr\'e $r$.
Notre implantation est $560$ fois plus rapide que l'algorithme na\"{\i}f
dans le cas $r = 24036583$ (exposant de Mersenne).

Avec notre programme,
nous avons trouv\'e deux nouveaux trin\^omes
primitifs de degr\'e $24036583$ sur $\GF(2)$, le pr\'ec\'edent record
\'etant de degr\'e $6972593$.
}
\RRabstract{
We give a new algorithm for performing the distinct-degree factorization of
a polynomial $P(x)$ over $\GF(2)$, using a multi-level
blocking strategy. The coarsest level of blocking replaces GCD computations
by multiplications, as suggested by Pollard (1975), von~zur Gathen and Shoup
(1992), and others.
The novelty of our approach is that
a finer level of blocking replaces multiplications by squarings,
which speeds up the computation
in $\GF(2)[x]/P(x)$ of certain \emph{interval polynomials}
when $P(x)$ is sparse.

As an application we give a fast
algorithm to search for all irreducible trinomials 
$x^r + x^s + 1$ of degree
$r$ over $\GF(2)$, while producing a certificate that can be checked 
in less time than the full search.
Naive algorithms cost $O(r^2)$ per trinomial, thus $O(r^3)$ to search
over all trinomials of given degree~$r$.
Under a plausible assumption about the 
distribution of factors of trinomials, the new algorithm has
complexity $O(r^2 (\log r)^{3/2}(\log\log r)^{1/2})$
for the search over all trinomials of degree ~$r$.
Our implementation achieves a speedup of greater than a factor
of $560$ over the naive algorithm 
in the case $r = 24036583$ (a Mersenne exponent).

Using our program, we have found two new primitive trinomials
of degree $24036583$ over $\GF(2)$ (the previous record degree was $6972593$).
}
\RRmotcle{
Complexit\'e amortie,
corps fini,
exposant de Mersenne,
factorisation en degr\'es distincts,
factorisation de polyn\^ome,
trin\^ome irr\'eductible,
trin\^ome primitif
}
\RRkeyword{
Amortized complexity, 
distinct degree factorization,
finite field,
irreducible trinomial, 
Mersenne exponent,
polynomial factorization, 
primitive trinomial}
\RRprojet{CACAO}
\RRtheme{\THSym} 
\URLorraine 
\begin{document}
\makeRR   

%
%
%
%


\title%
{A Multi-level Blocking Distinct Degree\\ Factorization Algorithm}

\maketitle

\section{Introduction}

The problem of factoring a univariate polynomial $P(x)$ over a finite
field $F$ often arises in computational 
algebra~\cite{Cantor92,Gathen02,Gathen92}.
An important case is 
when $F$ has small characteristic and $P(x)$ has high degree but is 
{\em sparse}, that is $P(x)$ has only a small number of nonzero terms.

To simplify the exposition we
restrict attention to the case where
$F = \GF(2)$ and $P(x)$ is a {\em trinomial}
\[P(x) = x^r + x^s + 1, \;\; r > s > 0, \]
although the ideas apply more generally and should be useful for factoring
sparse polynomials over fields of small characteristic.

Our aim is to give an algorithm with good {\em amortized complexity}, that
is, one that works well {\em on average}. Since we are restricting
attention to trinomials, we average over all trinomials of fixed degree~$r$.

Our motivation is to speed up previous algorithms for searching for
irreducible trinomials of high degree~\cite{rpb199,rpb214,Kumada00}.
For given degree $r$, we want to find all irreducible trinomials
$x^r + x^s + 1$. 

In our examples the degree $r$ is a {\em Mersenne exponent}, i.e., $2^r-1$ is
a Mersenne prime. In this case an irreducible trinomial of degree~$r$ is
necessarily primitive.  In general, without the restriction to Mersenne
exponents, we would need the prime factorisation of $2^r-1$ in order to test
primitivity (see e.g., 
\cite{Gathen99}). 

We are only interested in Mersenne exponents $r = \pm 1 \bmod 8$, because
in other cases Swan's theorem~\cite{Pellet78,Stickelberger97,Swan62} 
rules out irreducible trinomials of degree~$r$ (except for $s = 2$ or $r-2$, 
but these cases are usually easy to handle:
for example if $r = 13466917$ or $20996011$
we have $r = 1 \bmod 3$,
so $x^r + x^2 + 1$ is divisible by $x^2 + x + 1$).

Mersenne exponents can be found on the GIMPS website~\cite{GIMPS}. At the
time of writing, the five largest known Mersenne exponents $r$ satisfying
the condition $r = \pm 1 \bmod 8$ are $r = 6972593$, $24036583$, $25964951$,
$30402457$ and $32582657$. In the smallest case $r = 6972593$, a primitive
trinomial was found by Brent, Larvala and Zimmermann~\cite{rpb214} using an
efficient implementation of the naive algorithm.  However, it was not
feasible to consider the larger Mersenne exponents $r$ 
using the same algorithm,
since the time complexity of this algorithm is roughly of order $r^3$,
and the next case $r = 24036583$ would take about 41 times longer than
$r = 6972593$.
With the new ``fast'' algorithm described in this paper we have been able to
find two primitive trinomials of degree $r = 24036583$ in less time than
the naive algorithm took for $r = 6972593$. The speedup over the
naive algorithm for $r = 24036583$ is about a factor of $560$.

If $x^r + x^s + 1$ is reducible then we want to provide an
easily-checked {\em certificate} of reducibility. The certificate can simply
be an encoding of an irreducible factor $f$ of $x^r + x^s + 1$.
We choose the factor $f$ of smallest
degree $d > 0$. In case there are several factors of equal smallest degree
$d$, we give the one that is least in lexicographic order,
e.g., $x^3 + x + 1$ is preferred to $x^3 + x^2 + 1$.

\subsection{Distinct degree factorization}

Our basic algorithm performs 
{\em distinct degree factorization}~\cite{Flajolet01,Gathen99,Gathen02}.
That is, if $P(x)$ has several factors of the same degree $d$, the algorithm
will produce the product of these factors. The Cantor-Zassenhaus algorithm
is used to split this product into distinct factors. This is cheap
because the product usually consists of just one irreducible factor or is a
product of irreducible factors of small (equal) degree.

In the complexity analysis we only consider the time required to find {\em
one} nontrivial factor (it will be a factor of smallest degree) or output
``irreducible'', since that is what is required in the search for
irreducible trinomials.

\subsection{Factorization over $\GF(2)$}

It is well-known that $x^{2^d}+x$ 
is the product of all irreducible polynomials of degree dividing~$d$.
For example,
\[x^{2^3} + x = x(x+1)(x^3 + x + 1)(x^3 + x^2 + 1)\enddot\]
Thus, a simple algorithm to find a factor of smallest degree of $P(x)$ is
to compute $\GCD(x^{2^d} + x, P(x))$ for $d = 1, 2, \ldots$
The first time that the GCD is nontrivial, it contains a factor of minimal
degree~$d$. If the GCD has degree $>d$, it must be a product of factors
of degree~$d$.
If no factor has been found for $d \le r/2$, where $r = \deg(P(x)$), then
$P(x)$ must be irreducible.

Some simplifications are possible when $P(x) = x^r + x^s + 1$ is a
trinomial over $\GF(2)$ with $r$ or $s$ odd (otherwise $P(x)$ is trivially
reducible):

\begin{enumerate}
\item We can skip the case $d=1$ because a trinomial can not have
a factor of degree~$1$.
\item Since $x^rP(1/x) = x^r + x^{r-s} + 1$, 
we only need consider $s \le r/2$.
\item We can assume that $P(x)$ is square-free. 
\item By applying Swan's theorem, we can often show that the trinomial under
consideration has an odd number of irreducible factors; in this case we only
need check $d \le r/3$ before claiming that $P(x)$ is irreducible.
\end{enumerate}

\section{Complexity of the algorithm}

Note that $x^{2^d}$ should not be computed explicitly; 
it is much better to compute
$x^{2^d} \bmod P(x)$ by repeated squaring. The complexity of squaring
modulo a trinomial of degree $r$ is only $S(r) = O(r)$ bit-operations.

\subsection{Complexity of polynomial multiplication and squaring}

As well as performing GCD computations we need to perform multiplications in
$\GF(2)[x]/P(x)$, and an important special case is squaring a polynomial
modulo $P(x)$, so we first consider the bit-complexity of these operations.

Multiplication of polynomials of
degree $r$ over $\GF(2)$ can be performed in time
$M(r) = O(r \log r \log\log r)$. We have implemented an algorithm of
Sch\"onhage~\cite{Schonhage77} that achieves this bound. The algorithm uses
a radix-$3$ FFT and is different from the better-known
Sch\"onhage-Strassen algorithm~\cite{SS71}. We remark that the $\log\log r$
term in the time-bound for the Sch\"onhage-Strassen algorithm has been
reduced by F\"urer~\cite{Furer07}, but it is not clear if a similar idea
can be used to improve Sch\"onhage's algorithm~\cite{Schonhage77}.
In any event the 
$\log\log r$ term comes from the number of levels of recursion and is
a small constant for the values of~$r$ that we are considering.

In practice, Sch\"onhage's algorithm is not the fastest unless $r$ is quite
large. We have also implemented classical, Karatsuba and Toom-Cook
algorithms that have $M(r) = O(r^\alpha)$, $1 < \alpha \le 2$, since these
algorithms are easier to implement and are faster for small~$r$. Our
implementations of the Toom-Cook algorithms TC3 and TC4 are based on recent
ideas of Bodrato~\cite{Bodrato07}.

For brevity we assume that $r$ is
large and Sch\"onhage's algorithm is used. 
On a 64-bit machine the crossover versus TC4 occurs near degree $r = 108000$.

In the complexity estimates we assume that $M(r)$ is a sufficiently smooth
and well-behaved function.

By {\em Squaring} we mean squaring a polynomial of degree $<r$ and reduction
mod $P(x)$.  Squaring in $\GF(2)[x]/P(x)$ can be performed in time 
$S(r) = \Theta(r) \ll M(r)$ (assuming, as usual, that $P(x)$ is a trinomial).
Our algorithm takes advantage of the fact that squaring is much faster than
multiplication.

Where possible we use the memory-efficient squaring algorithm of Brent,
Larvala and Zimmermann~\cite{rpb199}, which in our implementation
is about $2.2$ times faster than the naive squaring algorithm.

\subsection{Complexity of GCD}

For GCDs we use a sub-quadratic algorithm that runs in time
$G(r) = \Theta(M(r)\log r)$.
More precisely,
\[G(2r) = 2G(r) + \Theta(M(r))\endcomma\] 
so for $\alpha > 1$,
\[M(r) = \Theta(r^\alpha) \Rightarrow G(r) = \Theta(M(r))\endcomma\] 
and
\[M(r) = \Theta(r \log r \log\log r) \Rightarrow
  G(r) = \Theta(M(r) \log r)\enddot\] 
In practice, for $r \approx 2.4\times 10^7$ and 
our implementation on a 2.2~Ghz Opteron,
$S(r) \approx 0.005$ second,
$M(r) \approx 2$ seconds,
$G(r) \approx 80$ seconds,
so $M(r)/S(r) \approx 400$,
and $G(r)/M(r) \approx 40$.

\subsection{Avoiding GCD computations}

In the context of integer factorization, Pollard~\cite{Pollard75} 
suggested a blocking
strategy to avoid most GCD computations and thus reduce the amortized cost;
von zur Gathen and Shoup~\cite{Gathen92} applied the same idea to polynomial 
factorization.

The idea of blocking is to choose a parameter $\ell > 0$ and, 
instead of computing
\[\GCD(x^{2^d} + x, P(x)) \;\;\mbox{for}\;\; d \in [d', d'+\ell)\endcomma\] 
compute
\[\GCD(p_\ell(x^{2^{d'}}, x), P(x))\endcomma\] 
where the {\em interval polynomial}
$p_\ell(X,x)$ is defined by
\[p_{\ell}(X,x) = \prod_{j=0}^{\ell-1} \left( X^{2^j} + x \right)\enddot\]
In this way we replace $\ell$ GCDs by one GCD and $\ell-1$ multiplications
mod $P(x)$. 

The drawback of blocking is that we may have to backtrack if $P(x)$ has
more than one factor with degree in the interval $[d', d'+\ell)$, 
since the algorithm produces the product of these factors.
Thus $\ell$ should not be too
large. The optimal strategy depends on the expected size distribution of
factors and the ratio of times for GCDs and multiplications.

\subsection{Multi-level blocking}

Our (apparently new) idea is to use
a finer level of blocking to replace most multiplications by squarings,
which speeds up the computation
in $\GF(2)[x]/P(x)$ of the above interval polynomials.
The idea is to split the interval $[d', d'+\ell)$ into $k \geq 2$ smaller
intervals of length $m$ over which
\begin{equation} \label{pofm}
p_{m}(X,x) = \prod_{j=0}^{m-1} \left( X^{2^j} + x \right)
= \sum_{j=0}^m x^{m-j} s_{j,m}(X)\endcomma
\end{equation}
where
\begin{equation}
s_{j,m}(X) = \sum_{0 \le k < 2^m,\; w(k) = j} X^k\endcomma		\label{eq:s}
\end{equation}
and $w(k)$ denotes the {\em Hamming weight} of $k$,
that is the number of nonzero bits in the binary representation of~$k$.

For example, for $m=3$, we have:
\[ p_m(X,x) = x^3 + x^2 (X^4 + X^2 + X) + x (X^6 + X^5 + X^3) + X^7, \]
where $s_{0,3}(X) = 1$, $s_{1,3}(X) = X^4 + X^2 + X$,
$s_{2,3}(X) = X^6 + X^5 + X^3$, and $s_{3,3}(X) = X^7$.

Note that 
\[s_{j,m}(X^2) = s_{j,m}(X)^2 \;\;\mbox{in}\;\; \GF(2)[x]/P(x)\enddot\]
Thus,
$p_{m}(x^{2^d},x)$
can be computed with cost $m^2S(r)$ if we already know
$s_{j,m}(x^{2^{d-m}})$ for $0 < j \le m$.
(The constant polynomial $s_{0,m}(X)=1$ is computed only once.)

Continuing the example with 
$m=3$, and assuming that we know $s_{1,3}(x^{2^{d-3}})$,
$s_{2,3}(x^{2^{d-3}})$, and $s_{3,3}(x^{2^{d-3}})$,
squaring each of these $m=3$ times gives
$s_{1,3}(x^{2^d})$, $s_{2,3}(x^{2^d})$, and $s_{3,3}(x^{2^d})$,
from which we can easily get $p_3(x^{2^d},x)$ using the sum in
Eq.~(\ref{pofm}).

In this way we replace $m-1$ multiplications and $m$ squarings~--- 
if we used the product in Eq.~(\ref{pofm})~---
by $m^2$ squarings.
Each $s_{j,m}$, $0 < j \le m$, requires $m$ squarings to be shifted from
argument $x^{2^{d-m}}$ to argument $x^{2^d}$.
The summation in Eq.~(\ref{pofm}) costs only $O(mr)$, which is negligible.
Choosing $m \approx \sqrt{M(r)/S(r)}$
(about $20$ if
$M(r)/S(r) \approx 400$), the speedup over single-level blocking is about 
$m/2 \approx 10$ (not counting the cost of GCDs).

Von zur Gathen and Gerhard~\cite[p.~1685]{Gathen02} suggested using the
same idea with $m=2$ (thus reducing the number of multiplications by a
factor of two), but did not consider choosing an optimal $m > 2$.

At first sight initialization of the polynomials $s_{j,m}(X)$ for $X = x$
might appear to be expensive, since the
definition~(\ref{eq:s}) involves $O(2^m)$ terms.
However, the polynomials $s_{j,m}(X)$
satisfy a ``Pascal triangle'' recurrence relation
\[ s_{j,m}(X) = s_{j,m-1}(X^2) + X s_{j-1,m-1}(X^2)\]
with boundary conditions
\[s_{j,m}(X) = \left\{ \begin{array}{ll}
		0 		&\mbox{if $j > m \ge 0$,}\\
		1 		&\mbox{if $m \ge j = 0$.}
		\end{array} \right.
\]
Using this recurrence,
it is easy to compute $s_{j,m}(x) \bmod P(x)$ for $0 \leq j \leq m$
in time $O(m^2r)$. Thus, the initialization is cheap.

To summarise, we use two levels of blocking:  

\begin{enumerate}
\item 
The outer level replaces most GCDs by multiplications.
\item 
The inner level replaces most multiplications by squarings.
\item 
The parameter $m \approx \sqrt{M(r)/S(r)}$ is used for 
the inner level of blocking.  
\item 
A different parameter $\ell = km$ is used
for the outer level of blocking.
\end{enumerate}

For example, suppose 
$S = 1/400$, $M = 1$, $G = 40$ (where we have normalised so $M = 1$).
We could choose $\ell = 80$ and $m = 20$.
With no blocking, the cost for an interval of length $80$ is
$80G + 80S = 3200.2$;
with 1-level blocking the cost is $G + 79M + 80S = 119.2$;
with 2-level blocking the cost is $G + 3M + 1600S = 47.0$.

\subsection{Sieving out small factors} \label{sieving}

We define a
{\em small} factor to be one with degree $d < \frac{1}{2}\log_2 r$, 
so $2^d < \sqrt{r}$. The constant $\frac{1}{2}$ in the definition is
arbitrary and could be replaced by any fixed constant in $(0,1)$.
A {\em large} factor is a factor that is not small.

It would be inefficient to find small factors in the same way as large
factors.  
Instead, let $D = 2^d-1$, $r' = r \bmod D$, $s' = s \bmod D$.
Then
\[P(x) = x^r + x^s + 1 = x^{r'} + x^{s'} + 1 \bmod (x^{D} - 1)\endcomma\]
so we only need compute
\[\GCD(x^{r'} + x^{s'} + 1, x^{D} - 1)\enddot\]
Because $r', s' < D < \sqrt{r}$, the cost of finding 
small factors is negligible (both theoretically and in practice),
so can be neglected.

\subsection{Outer level blocking strategy} \label{subsec:outer}
 
The blocksize in the outer level of blocking is $\ell = km$. We take
a linearly increasing sequence of block sizes
\[k = k_0 j \;\;\mbox{for}\;\; j = 1, 2, 3, \dots\endcomma\]
where the first interval starts at about 
$\log r$ (since small factors will have
been found by sieving).
 
The choice $k = k_0 j$ leads to a quadratic polynomial for the interval
bounds;
other
possibilities 
are discussed by von zur Gathen
and Gerhard~\cite{Gathen02}.

In principle, using the data that we have obtained on the distribution of
degrees of smallest factors of trinomials (see \S\ref{sec:dist}), 
and assuming that this
distribution is not very sensitive to the degree $r$, we could obtain a
strategy that is close to optimal.  However, the choice
$k_0 j$ with suitable $k_0$ is easy to implement and not too far from 
optimal. The number of GCD and sqr/mul operations is usually 
within a factor of $1.5$ of the minimum possible 
in our experiments.

\section{Distribution of degrees of factors}		\label{sec:dist}

In order to predict the expected behaviour of our algorithm, we need
to know
the expected distribution of degrees of smallest irreducible factors. 
{From} Swan's theorem~\cite{Swan62}, we know that there are
significant differences between the distribution of factors of trinomials
and of all polynomials of the same degree.
Our complexity estimates are based on the heuristic assumption that this 
difference is not too large, in a sense made precise by
Hypothesis~\ref{hyp1}.
\begin{hypothesis} 					\label{hyp1}
Over all trinomials $x^r+x^s+1$ of degree $r$ over $\GF(2)$,
the probability $\pi_d$ that a trinomial has no nontrivial factor of degree
$\le d$, $1 < d \le r$, is at most $c/d$, where $c$ is a constant.
\end{hypothesis}

Hypothesis~\ref{hyp1} implies that there are at most $c$ irreducible
trinomials of degree~$r$. This is probably false, as there may well be
a sequence of exceptional $r$ for which the number of irreducible trinomials
is unbounded. Thus, we may need to replace the constant $c$ in
Hypothesis~\ref{hyp1} by a slowly-growing function $c(r)$.
Nevertheless, in order to give realistic complexity estimates that are
in agreement with experiments, we assume below that Hypothesis~\ref{hyp1}
is correct. Under this assumption
we use an amortized model to obtain
the total complexity over all trinomials of degree $r$.

{From} Hypothesis~\ref{hyp1}, the probability that a trinomial does not
have a small factor (as defined in \textsection\ref{sieving}) is $O(1/\log r)$.

Table~\ref{table:3M24M} gives the observed values of $d\pi_d$ for
$r=3021377$, $r=6972593$, and  $r=24036583$. The maximum values for
each $r$ are given in bold. 
The table shows that the values of $d\pi_d$
are remarkably stable for small~$d$, and bounded by~$4$ for large~$d$
(this is because there are four irreducible trinomials of degree $3021377$
and also four of degree $24036583$, when we count 
both trinomials $x^r + x^s + 1$ and
their reciprocals $x^r + x^{r-s} + 1$).

\begin{table}[ht]			
\centering
\caption{$d\pi_d$ for various degrees $r$.} 
							\label{table:3M24M}
\begin{tabular}{|c|c|c|c|}
\hline
$d$	& $r=3021377$	& $r=6972593$	& $r=24036583$	\\ 
\hline
2	& 1.333		& 1.333		& 1.333\\
3	& 1.429		& 1.429		& 1.429\\
4	& 1.524		& 1.524		& 1.524\\
5	& 1.536		& 1.536		& 1.536\\
6	& 1.598		& 1.598		& 1.598\\
7	& 1.600		& 1.600		& 1.600\\
8	& 1.667		& 1.667		& 1.667\\
9	& 1.642		& 1.642		& 1.642\\
10	& 1.652		& 1.652		& 1.652\\
100	& 1.763		& 1.771		& 1.770\\
1000	& 1.783		& 1.756		& 1.786\\
10000	& 1.946		& 1.873		& 1.786\\
100000	& 1.986		& 1.606		& 1.880\\
279383	& 1.480		& {\bf 2.084}	& 1.813\\
1000000 & 1.324		& 1.147		& 1.831\\
10000000& --		& --		& 1.664\\
$r-1$	& {\bf 4.000}	& 2.000		& {\bf 4.000}\\
\hline
\end{tabular}
\end{table}

\subsection{Consequences of the hypothesis}

Define $p_k = \pi_{d-1} - \pi_d$ to be the
probability that the smallest nontrivial factor $f$ of a randomly
chosen trinomial has degree $d = \deg(f)$.  In order to estimate the
running time of our algorithm, we use the following Lemma, which gives
the expectation $E_\beta$ of $d^\beta$.

\begin{lemma}						\label{expectA}
If $\beta > 0$ is constant and Hypothesis~\ref{hyp1} holds,
then
\[E_{\beta} := \sum_{d=1}^r d^\beta p_d = 
	\left\{ \begin{array}{ll}
	O(1)		&\mbox{if $\beta < 1$,}\\
	O(\log r)	&\mbox{if $\beta = 1$,}\\
	O(r^{\beta-1})	&\mbox{if $\beta > 1$.}
	\end{array} \right.
\]
\end{lemma}
\begin{proof}
We use summation by parts.
Note that a trinomial has no factor of degree~$1$, so $p_1 = 0$
and $\pi_0 = \pi_1 = 1$. Thus
\begin{eqnarray*}
E_\beta &=& \sum_{d=1}^r d^\beta p_d 
	 \;\;=\;\; \sum_{d=1}^r d^\beta (\pi_{d-1} - \pi_d) \\
	 &=& \sum_{d=1}^{r-1} \left((d+1)^\beta - d^\beta\right)\pi_d
		+ \pi_0 - r^\beta\pi_r \\
	 &\le& 1 + c\sum_{d=1}^{r-1} \frac{(d+1)^\beta - d^\beta}{d}
		\;\;\mbox{(by Hypothesis~\ref{hyp1})} \\
         &\le& 1 + O\left(\sum_{d=1}^{r-1} d^{\beta-2}\right)
\end{eqnarray*}
and the result follows.
\end{proof}

The following Lemma gives a stronger result in the case $\beta < 1$.

\begin{lemma}						\label{expectB}
If $0 < \beta < 1$, $0 < D \le r$, and Hypothesis~\ref{hyp1} holds,
then
\[\sum_{d=D}^r d^\beta p_d = O\left(D^{\beta-1}\right)\enddot
\]
\end{lemma}
\begin{proof}
The proof is similar to that of Lemma~\ref{expectA}. We end with the
upper bound
\[
\sum_{d=D}^{r-1}\frac{(d+1)^\beta - d^\beta}{d} + D^{\beta}\pi_{D-1}
\enddot\]
{From} Hypothesis~\ref{hyp1}, 
$\pi_{D-1} = O(1/D)$, and the sum over $d$ is $O(D^{\beta-1})$,
so the result follows.
\end{proof}

\section{Expected cost of sqr/mul and GCD}

Recall that the inner level of blocking replaces $m$ multiplications by
$m^2$ squarings and one multiplication, where the choice
$m \approx \sqrt{M(r)/S(r)}$ makes the total cost of squarings about equal to
the cost of multiplications.

For a smallest factor of degree $d$, the number of squarings is
$m(d + O(\sqrt{d}))$,
where the $O(\sqrt{d})$ term follows from our choice of outer-level
blocksizes (see~\S\ref{subsec:outer}).
Averaging over all trinomials of degree~$r$,
the expected number of squarings is
\[O\left(m\;\sum_{d \le r/2} (d + O(\sqrt{d}))p_d\right)\endcomma\]
and from Lemma~\ref{expectA} this is 
$O(m\log r)$.
Thus, the expected cost of sqr/mul operations per trinomial is
\begin{eqnarray}
O\left(S(r)\log r \sqrt{M(r)/S(r)}\right)
&=& O\left(\log r \sqrt{M(r)S(r)}\right)		    \nonumber \\
&=& O\left(r (\log r)^{3/2}(\log\log r)^{1/2}\right)\enddot \label{eq:S1}
\end{eqnarray}
If we used only a single level of blocking, then
the cost of multiplications would dominate that of 
squarings, with an expected cost per trinomial of 
$O\left(\log r M(r) \right) = O\left(r (\log r)^2 \log\log r \right)$.

(\ref{eq:S1}) is correct as $r \to \infty$. However, in practice,
at least for $r < 6.4 \times 10^7$,
our implementation of Sch\"onhage's FFT-based polynomial multiplication 
algorithm~\cite{Schonhage77} calls a different multiplication
routine (usually TC4) to perform smaller multiplications, 
rather than recursively calling itself. TC4 has 
exponent $\alpha' = \ln(7)/\ln(4) \approx 1.4$, so the effective
exponent for FFT multiplication is 
$\alpha = (1 + \alpha')/2 \approx 1.2 > 1$.  In this case, the
expected cost of sqr/mul operations per trinomial is
\begin{equation}
O\left(\log r \sqrt{M(r)S(r)}\right) =
O(r^{(1+\alpha)/2}\log r) = O(r^{1.1\cdots}\log r)	\label{eq:S2}
\end{equation}

\subsection{Expected cost of GCDs}

Suppose that $P(x)$ has a smallest factor of degree~$d$. The number of GCDs
required to find the factor, using our (quadratic polynomial) 
blocking strategy, 
is at least $1$, and $O(\sqrt{d})$ if $d$ is large. By Hypothesis~\ref{hyp1},
the expected number of GCDs for a trinomial with {\em no small factor} is
\[1 + O\left(\sum_{\log_2 r < 2d \le r} d^{1/2}\;p_d \right)\endcomma\]
and by Lemma~\ref{expectB} this is
\[1 + O\left(\frac{1}{\sqrt{\log r}}\right)\enddot\] 
Thus the expected cost of GCDs per trinomial is
\begin{equation} 
O(G(r)/\log r) = O(M(r)) = O(r\log r \log\log r)\enddot		\label{eq:G1}
\end{equation}
(\ref{eq:G1}) is asymptotically less than the expected cost 
(\ref{eq:S1}) of sqr/mul operations.
However, 
if $M(r) = O(r^\alpha)$ with $\alpha > 1$, 
then the expected cost of GCDs is $O(r^\alpha/\log r)$, which is
asymptotically greater than the expected cost~(\ref{eq:S2})
of sqr/mul operations.
Note the expected cost of GCDs does not depend on whether we use
one or two levels of blocking.

For $r \approx 2.4 \times 10^7$,
GCDs take about 65\% of the time versus 35\% for sqr/mul.

\subsection{Comparison with previous algorithms}

For simplicity we use the $\Otilde$ notation which ignores $\log$
factors. For example, $M(r) = \Otilde(r)$.

The ``naive'' algorithm, as implemented by Brent, Larvala and
Zimmermann~\cite{rpb199} and earlier authors, takes an expected time
$\Otilde(r^2)$ per trinomial, or $\Otilde(r^3)$ to cover all trinomials of
degree~$r$.

The single-level blocking strategy and the new algorithm both take
expected time $\Otilde(r)$ per trinomial,
or $\Otilde(r^2)$ to cover all trinomials of degree~$r$.

In practice, the new algorithm is faster over the naive algorithm
by a factor of about $160$ for
$r = 6972593$, and by a factor of about $560$ for $r = 24036583$.
For $r = 24036583$, where
sqr/mul operations take 35\% of the total time in the new algorithm,
and the corresponding speedup is about 10, this gives a global
speedup of more than 4 over the single-blocking strategy.

\subsection{Some details of our implementation}

We first implemented the\linebreak 
2-level blocking strategy in NTL \cite{NTL}.
To get full efficiency, we rewrote all critical routines
and tuned them efficiently on the target processors.
Our squaring routine implements the algorithm described in \cite{rpb199},
which is more than twice as fast as the corresponding optimized
NTL routine for trinomials.
Our multiplication routine implements Toom-Cook $3$-way, $4$-way, and
Sch\"onhage's algorithm \cite{Schonhage77}.
We also improved the basecase multiplication code; more details 
concerning
efficient multiplication in $\GF(2)[x]$ will be published in \cite{BrGaThZi07}.
Finally, we implemented a subquadratic GCD routine, since NTL only
provides a classical GCD for binary polynomials.

\subsection{Primitive trinomials}


The largest published primitive trinomial is 
\[x^{6972593} + x^{3037958} + 1\endcomma\]			
found by Brent, Larvala and Zimmermann~\cite{rpb199} 
in 2002 using a naive (but efficiently implemented) algorithm.

In March--April 2007, we tested our new program by verifying 
the published results on primitive 
trinomials for Mersenne exponents $r \le 6972593$, and in the process produced
certificates of reducibility 
(lists of smallest factors for each reducible trinomial). These are 
available from the first author's website~\cite{rpbweb}.

In April--August 2007, we ran our new algorithm to search for
primitive trinomials of degree $r = 24036583$. This is the next Mersenne
exponent, apart from two that are trivial to exclude by Swan's theorem.
It would take about 41 
times as long as for $r = 6972593$ by the naive algorithm,
but our new program is 560 times
faster than the naive algorithm. 
Each trinomial takes on average about 16 seconds
on a 2.2~Ghz Opteron.

The complete computation was performed in four months, using
about 24 Opteron and Core~2 processors located
at ANU 
and INRIA. 

We found two new primitive trinomials of (equal) record degree:
\begin{equation}
x^{24036583} + x^{8412642} + 1		\label{Eugenie}   
\end{equation}
and
\begin{equation}
x^{24036583} + x^{8785528} + 1\enddot	\label{Judy-anne} 
\end{equation}

\subsection{Verification}

Allan Steel~\cite{Steel} kindly verified irreducibility of
(\ref{Eugenie})--(\ref{Judy-anne}) using Magma~\cite{Magma}.
Each verification took 
about 67 hours on an 2.4~GHz 
Core~2 processor.
Independent verifications using our {\tt irred V3.15}
program~\cite{rpb199,rpb214}
took about 35 hours on a 2.2~Ghz Opteron.  
The difference in speed is mainly due to the fast squaring algorithm
implemented in {\tt irred}.

Primitivity of (\ref{Eugenie})--(\ref{Judy-anne}) follows from irreducibility
provided that the degree $24036583$ is a Mersenne exponent. We have not 
verified this, but rely on computations performed by the GIMPS 
project~\cite{GIMPS}.

Reducibility of the remaining trinomials of degree $24036583$ can be
verified using the certificate (or {\em extended log}, 
a list of smallest irreducible factors) available from our
website~\cite{rpbweb}.
The verification takes less than $10$ hours using Magma on a 2.66~Ghz Core~2
processor.	

\section{Conclusion}

The new double-blocking strategy, combined with fast
multiplication and GCD algorithms, has allowed us to find new primitive
trinomials of record degree. 

The same ideas should work over finite fields $\GF(p)$ for small prime $p > 2$,
and for factoring sparse polynomials $P(x)$ that are 
not necessarily trinomials:
all we need is that the time for $p$-th powers (mod $P(x)$) 
is much less than the time for multiplication (mod $P(x)$).

\subsection*{Acknowledgements}

We thank Allan Steel for verifying irreducibility of the 
trinomials~(\ref{Eugenie})--(\ref{Judy-anne}),
and Marco Bodrato, Pierrick Gaudry and Emmanuel Thom\'e for 
their assistance in implementing fast algorithms for multiplication
of polynomials over $\GF[2]$.
ANU and INRIA provided computing facilities. The first author's 
research was supported by MASCOS and the Australian Research Council.

\bibliographystyle{amsalpha}


\end{document}